# Hydrogen Bonding to the Cysteine Ligand of Superoxide Reductase. Acid-Base Control of the Reaction Intermediates

*Emilie Tremey [1,2,3], Florence Bonnot [1,2,3], Yohann Moreau [1,2,3], Catherine Berthomieu [4], Alain Desbois [5], Vincent Favaudon [6], Geneviève Blondin [1,2,3], Chantal Houée-Levin [7], and Vincent Nivière [1,2,3*]*

[1] Laboratoire de Chimie et Biologie des Métaux, CEA, iRTSV, 17 avenue des Martyrs, Grenoble F-38054, France. [2] CNRS, UMR 5249, Grenoble, F-38054, France. [3] Université de Grenoble, Grenoble, F-38000, France. [4] Laboratoire des Interactions Protéine Métal, SBVME-CEA Cadarache/CNRS/Université Aix-Marseille II, 13108 Saint-Paul-lez-Durance Cedex, France. [5] Laboratoire Stress Oxydant et Détoxication, SB$^2$SM and CNRS UMR 8221, iBiTec-S, CEA Saclay, 91191 Gif-sur-Yvette Cedex, France. [6] Inserm Unité 612 and Institut Curie, Bâtiment 110-112, Centre Universitaire 91405 Orsay Cedex, France. [7] Laboratoire de Chimie Physique, CNRS/Université Paris-Sud, Bâtiment 350, Centre Universitaire 91405 Orsay Cedex, France.

Address correspondence to V. Nivière, Tel.: 33-4-38-78-91-09; Fax: 33-4-38-78-91-24;
E-mail: vniviere@cea.fr

*Abbreviations.* SOD, superoxide dismutase; SOR, superoxide reductase; FTIR, Fourier transform infrared; RR, resonance Raman; EPR, electron paramagnetic resonance; DFT, density functional theory

ABSTRACT

Superoxide reductase SOR is a non-heme iron metalloenzyme that detoxifies superoxide radical in microorganisms. Its active site consists of an unusual non-heme $Fe^{2+}$ center in a [$His_4$ $Cys_1$] square pyramidal pentacoordination, with the axial cysteine ligand proposed to be an essential feature in catalysis. Two NH peptide groups from isoleucine 118 and histidine 119 establish H-bondings with the sulfur ligand (*Desulfoarculus baarsii* SOR numbering). In order to investigate the catalytic role of these H-bonds, the isoleucine 118 residue of the SOR from *Desulfoarculus baarsii* was mutated into alanine, aspartate or serine residues. Resonance Raman spectroscopy showed that the mutations specifically induced an increase of the strength of the $Fe^{3+}$-S(Cys) and S-$C_\beta$(Cys) bonds as well as a change in conformation of the cysteinyl side chain, which was associated with the alteration of the NH hydrogen bonding to the sulfur ligand. The effects of the isoleucine mutations on the reactivity of SOR with $O_2^{\bullet-}$ were investigated by pulse radiolysis. These studies showed that the mutations induced a specific increase of the p$K_a$ of the first reaction intermediate, recently proposed to be an $Fe^{2+}$-$O_2^{\bullet-}$ species. These data were supported by DFT calculations carried out on three models of the $Fe^{2+}$-$O_2^{\bullet-}$ intermediate, with one, two or no H-bonds on the sulfur ligand. Our results demonstrated that the hydrogen bonds between the NH (peptide) and the cysteine ligand tightly control the rate of protonation of the $Fe^{2+}$-$O_2^{\bullet-}$ reaction intermediate to form an $Fe^{3+}$-OOH species.

## INTRODUCTION

Superoxide radical, $O_2^{\bullet-}$, the one-electron reduction product of oxygen, is a side product of aerobic metabolism [1, 2]. It is considered to be the first reactive oxygen species formed within the cells, initiating toxic oxidative stress processes which dramatically affect cellular metabolism and viability [1, 2]. It is now well documented that high levels of oxidative stress in human cells are involved in the development of serious diseases, like cancers or pathologies related with aging, Alzheimer or Parkinson diseases [3]. Detoxification of $O_2^{\bullet-}$ is thus a crucial part of the cellular antioxidant defense mechanisms which allow the cells to cope with the presence of molecular oxygen [1, 2]. So far, only two superoxide detoxification systems have been described. The first one is the well-known superoxide dismutase (SOD), present in almost all aerobic organisms, catalyzing dismutation of $O_2^{\bullet-}$ into $H_2O_2$ and $O_2$ [4]. The second one, superoxide reductase (SOR), has been discovered more recently [5, 6] and up to now has been only found in microorganisms [7-10]. SOD and SOR are structurally unrelated metalloproteins and carry out different reactions. SOR catalyzes the one-electron reduction of $O_2^{\bullet-}$ into $H_2O_2$ (eq. 1), the electron being provided by cellular reductases or soluble electron transfer proteins [7-10]:

$$O_2^{\bullet-} + 1e^- + 2H^+ \rightarrow H_2O_2 \qquad \text{(eq. 1)}$$

Although different classes of SOR have been described depending on the presence of an additional structural domain, which can bind a rubredoxin-like iron center (named center I), they all exhibit the same active site, named center II [7-10]. The SOR active site in its reduced form consists of an unusual non-heme $Fe^{2+}$ center in a [$His_4$ $Cys_1$] square pyramidal pentacoordination [11-14], which reacts at a nearly diffusion-controlled rate with $O_2^{\bullet-}$ [15-19]. The catalytic mechanism of SORs has been extensively studied, mainly using pulse radiolysis [15-19]. Ferric iron hydroperoxide [14-18] and more recently ferrous iron-superoxo species [19, 20] have been proposed as reaction intermediates during the course of superoxide reduction to produce $H_2O_2$. A first reaction intermediate T1 is rapidly formed upon the binding of $O_2^{\bullet-}$ to the vacant sixth coordination position of the $Fe^{2+}$ site (Scheme 1). In the case of the *Desulfoarculus baarsii* enzyme, T1 was proposed to be a ferrous iron-superoxo species [19]. This first reaction intermediate undergoes a protonation process by $H_3O^+$ from the bulk solvent, to form a second reaction intermediate, an $Fe^{3+}$-hydroperoxo species (T1' in Scheme 1). The second reaction intermediate was proposed to be specifically protonated on its proximal oxygen atom by a water molecule, hydrogen bonded to lysine 48, a highly conserved residue located close to the iron site, thus promoting the formation of the $H_2O_2$ product [14, 19] (Scheme 1). Although this second reaction intermediate could not be observed by pulse radiolysis, probably because it did not accumulate [19], its presence was deduced both from the X-ray structures of iron-peroxide intermediates trapped in the active site of SOR [14] and from the effects of the K48I mutation on the reactivity of SOR [19]. In particular, it was proposed that the mutation of lysine 48 into isoleucine, by preventing a specific protonation of the proximal oxygen of the $Fe^{3+}$-OOH species, favors its O-O bond cleavage to form a high valent iron-oxo species, which is avoided in the wild-type enzyme [19]. Finally, following the release of $H_2O_2$, the $Fe^{3+}$ active site remains hexacoordinated with the presence of $HO^-/H_2O$ ligand in the sixth coordination position (P1 in Scheme 1). The $HO^-/H_2O$ ligand is further replaced by the carboxylate side chain of the conserved glutamate 47 residue, to form a hexacoordinated $Fe^{3+}$-E47 species at the end of the catalytic cycle [17, 19] (P2 in Scheme 1).

The strictly conserved cysteine axial ligand in the SOR active site has been hypothesized to be an essential feature for the reactivity of the enzyme with superoxide [7-10]. However, cysteine substitution to probe its function is not possible, since it would lead, most likely, to the formation of a demetallated SOR active site [21]. Interestingly, several amino-acid residues were pointed out to interact with the sulfur atom of the cysteine ligand, suggesting that they could modulate the properties of the S-Fe bond and thus the reactivity of SOR with $O_2^{\bullet -}$. In *D. baarsii* SOR, the buried carboxylic side chain of E114 establishes a dipolar interaction with the sulfur ligand (C116) [22, 23] and its mutation into an alanine (E114A) was shown to specifically decrease the strength the Fe-S(Cys) bond [23]. Such a modification on the strength of the Fe-S(Cys) bond was reported to induce an increase of the p$K_a$ of the first reaction intermediate T1, providing experimental evidence for a role of the sulfur ligand in the control of the evolution of the reaction intermediate in SOR catalysis [23].

By comparison with other Fe-S(Cys) containing metalloproteins, cytochrome P450, rubredoxins or ferredoxins [24-26], it was proposed that in SORs, two NH peptide groups from leucine/isoleucine and histidine of a well conserved tetrapeptide C116-N117-L/I118-H119 are within H-bonding distance to the sulfur iron ligand [11-14, 27] (Figure 1). In the crystal structure of the SOR from *D. baarsii*, these N(H)-S distances are 3.4 Å and 3.2 Å for the I118 and H119 backbone, respectively (Figure 1), in agreement with the fact that H-bonds involving a sulfur atom are longer than those involving an oxygen atom. These N(H)-S bond distances in SOR are slightly shorter than those reported for the sulfur ligands of rubredoxin (3.6-3.9 Å) [26].

In this work, we have investigated the effect of one of the peptide NH groups on the sulfur ligand and on the reactivity of the SOR from *D. baarsii* with $O_2^{\bullet -}$ by mutating the I118 residue into A, D, or S. In these mutants, the steric constraints exerted by the lateral chains are expected to impact the positioning of the NH peptide group with respect to the sulfur ligand. The H119 position was not mutated since the histidine side-chain is directly involved in the iron coordination and its substitution, most likely, would result in important alterations of the active site.

Our data show that mutations of I118 induce a strengthening of the Fe-S(Cys) and S-$C_\beta$(Cys) bonds and a decrease of the p$K_a$ of the reaction intermediates. These results are consistent with the decrease of the strength of an hydrogen bond between the I118 peptide NH group and the sulfur ligand. These effects are opposite to those reported for the E114A mutant [23], showing that, in the SOR from *D. baarsii*, the strength of the Fe-S(Cys) bond is finely tuned by different second coordination sphere residues.

## MATERIAL AND METHODS

*Materials*. For pulse radiolysis experiments, sodium formate and buffers were of the highest grade available (Prolabo Normatom or Merck Suprapure). Oxygen was from ALPHA GAZ. Its purity is higher than 99.99%. Water was purified using an Elga Maxima system (resistivity 18.2 MΩ). $K_2IrCl_6$ was from Strem Chemical Inc.

*Site-directed mutagenesis and protein purification*. Mutagenesis was carried out on the plasmid pMJ25 [6], with the QuickChange® site-directed mutagenesis kit from Stratagene. Plasmid pMJ25 is a pJF119EH derivative, in which the *sor* gene from *D. baarsii* was cloned under the control of a *tac* IPTG inducible promotor [6]. Six primers were designed for the PCR-based site-directed mutagenesis to create the three I118 SOR mutants, I118A, I118S and I118D. Primer 1 (5' GAA TAC TGC AAC GCC CAC GGC CAC TGG 3') and primer 2 (5'CCA GTG GCC GTG GGC GTT GCA GTA TTC 3') were used for the I118A mutation. Primer 3 (5' GAA TAC TGC AAC GAC CAC GGC CAC TGG 3') and primer 4 (5' CCA GTG GCC GTG GTC GTT GCA GTA TTC 3') were used for the I118D mutation. Primer 5 (5' GAA TAC TGC AAC AGC CAC GGC CAC TGG 3') and primer 6 (5' CCA GTG GCC GTG GCT GTT GCA GTA TTC 3') were used for the I118S mutation. The mutations were verified by DNA sequencing. The resulting plasmids, pTE118A, pTE118S and pTE118D were transformed in an *E. coli* BL21 strain. Over-expressions and purifications of the I118 mutant proteins were carried out as reported for the wild-type protein [6].

The purified mutant proteins were found to be as stable as the wild-type protein. The UV-vis absorption spectrum of the purified proteins exhibited an $A_{280nm}/A_{503nm}$ ratio of 4.8-4.9 and appeared to be homogeneous, as judged by SDS-PAGE analysis (data not shown). Protein concentrations were determined using the Bio-Rad® protein assay reagent. For each SOR mutant, full metallation of the two mononuclear iron sites was verified by atomic absorption spectroscopy, with two iron atoms per polypeptide chain (data not shown). All the I118 mutant proteins were isolated with an oxidized center I ($\varepsilon_{503nm}$ = 4,400 $M^{-1}$ $cm^{-1}$) and a fully reduced center II. Fully oxidized SORs were obtained by treatment of the proteins with a slight molar excess of $K_2IrCl_6$.

*pH studies*. The following buffers were used to cover a pH range from 5.0 to 10.5: acetate for pH 5.0 and 5.5; MES for pH 5.6, 6.0 and 6.5; Bis-Tris propane or HEPES for pH 7.0; Tris-HCl for pH 7.6, 8.1, 8.5 and 8.8; glycine-NaOH for pH 9.1, 9.5, 10.2 and 10.5. The apparent p$K_a$ of the alkaline transition was determined from the pH dependence of the absorbance at 660 nm of the fully oxidized SOR mutant proteins, as reported in [16]. The titration curve was fitted with the equation expected from a single protonation process, $A_{660nm} = (A_{660max} + A_{660min} \times 10^{(pH-pKa\ app)})/(1 + 10^{(pH-pKa\ app)})$.

*Redox titrations*. The redox titrations were monitored by UV-visible absorption at 18 °C under an $N_2$ atmosphere in a Jacomex glove box equipped with a UV-visible cell coupled to an Uvikon XL spectrophotometer by optical fibers (Photonetics system). The sample cuvette contained 3 ml of 50 µM of protein solution in 50 mM Tris-HCl buffer, pH 7.6, in the presence of the following mediators at 2 µM each: ferrocene (+422 mV), N, N'-dimethyl-p-phenylenediamine (+371 mV), 1,4-benzoquinone (+280 mV), 2.5-dimethyl-p-benzoquinone (+180 mV), duroquinone (+5 mV). The redox potentials were measured with a combined Pt-Ag/AgCl/KCl (3 M) microelectrode, with respect to a standard hydrogen electrode. The potential of the solution was poised by stepwise additions of small quantities of either 5 mM of $K_2IrCl_6$ (starting from as-isolated SOR), or 10 mM sodium dithionite (starting from oxidized SOR). After each addition of $K_2IrCl_6$ or sodium dithionite, the solution was equilibrated during 5 min. After the potential reading had stabilized, the potential and the UV-visible absorption spectrum were recorded. The data were fitted to the Nernst equation for a one-electron reaction.

*Pulse radiolysis.* Pulse radiolysis measurements were performed as described elsewhere [28]. Briefly, free radicals were generated by irradiation of $O_2$-saturated aqueous protein solutions (100 μM), in 2 mM buffer, 10 mM sodium formate with 0.2-2 μs pulses of 4 MeV electrons at the linear accelerator at the Curie Institute, Orsay, France. Superoxide anion, $O_2^{\bullet -}$, was generated during the scavenging by formate of the radiolytically produced hydroxyl radical, $HO^{\bullet}$ [28]. The doses per pulse were calibrated from the absorption of the thiocyanate radical $(SCN)_2^{\bullet -}$ obtained by radiolysis of the thiocyanate ion in $N_2O^-$ saturated solution ($[SCN^-] = 10^{-2}$ M, $G((SCN)_2^{\bullet -}) = 0.55$ μmol. $J^{-1}$, $\varepsilon_{472\ nm} = 7580$ $M^{-1}$ $cm^{-1}$). For instance, a dose of 5 Gy per pulse (0.2 μs long) resulted in 2.8 μM of $O_2^{\bullet -}$. The dose varied linearly with the pulse length. Reactions were followed spectrophotometrically, using a Hamamatsu SuperQuiet xenon-mercury lamp (150 W) between 310 and 750 nm or a tungsten lamp between 450 and 750 nm, at 20 °C in a 2 cm path length fused silica cuvette. In all the pulse radiolysis experiments, a cut-off filter cutting all wavelengths below 425 nm was positioned between the lamp and the cuvette. Kinetics were analyzed at different wavelengths using a Levenberg-Marquardt algorithm from the Kaleidagraph® software package (Synergy Software).

*Resonance Raman spectroscopy.* For the Raman experiments, 3.5 μL of concentrated protein (4-6 mM) were deposited on a glass slide sample holder and then transferred into a cold helium gas circulating optical cryostat (STVP-100, Janis Research), held at 15 K. Resonance Raman spectra were recorded using a Jobin-Yvon U1000 spectrometer, equipped with a liquid nitrogen-cooled CCD detector (Spectrum One, Jobin-Yvon, France). Excitation at 647.1 nm (30 mW) was provided by an Innova $Kr^+$ laser (Coherent, Palo Alto). The signal-to-noise ratios were improved by spectral collections of 6 cycles of 30 s accumulation time. The reported spectra were the results of averaging 5-7 single spectra. The spectral analysis was made using the Grams 32 software (Galactic Industries). The spectrometer calibration was done as previously described in [29]. The frequencies of the resonance Raman bands of SORs were also internally calibrated against the main band of the ice lattice (230 $cm^{-1}$) and a residual laser emission at 676.4 nm (669 $cm^{-1}$). The frequency precision was 0.5-1 $cm^{-1}$ for the most intense bands and 1.5-2 $cm^{-1}$ for the weakest bands.

*FTIR*. Electrochemically-induced FTIR difference spectra were recorded, using a thin pathlength electrochemical cell and a Bruker 66 SX spectrometer equipped with a KBr beam splitter and a nitrogen-cooled MCT-A detector, as described in [22]. 10 µL of a 1 mM solution of SOR in a 50 mM Tris-HCl buffer pH 7.5, 20 mM $MgCl_2$ and 100 mM KCl were used for one sample. Ferrocene, N,N-dimethyl-p-phenylenediamine, 1,4-benzoquinone, N,N,N’,N’-tetra-methyl-p-phenylenediamine, 2,3,4,5-tetramethyl-p-phenylenediamine, phenazine ethosulfate, and duroquinone were used as electrochemical mediators, at a concentration of 40 µM.

*Electronic absorption and EPR spectroscopies.* Optical absorbance measurements were performed using a Varian Cary spectrophotometer (0.2 nm bandwidth) with 1 cm path length cuvette. Low-temperature (4.2 K) X-band EPR spectra were recorded on a Bruker EMX 081 spectrometer equipped with an Oxford Instrument continuous flow cold He gas cryostat.

*DFT calculations*. Three models (designed A, B and C) of the SOR active site have been built using the structure of the wild-type SOR with the 2IJ1 PDB code [14]. All include the side chains plus the Cα atoms of the five amino-acids binding the iron atom (residues H49, H69, H75, H119, C116). In model A, α-carbons were kept fixed in their crystallographic positions, while the hydrogens bound to them (replacing the adjacent atoms) were free to move only along the Cα-N or Cα-C direction found in the crystallographic structure. Model B is similar to A except that the peptide bond between H119 and I118 was added (i.e. a –NH-COH moiety replaces the hydrogen atom of model A). In order to maintain the orientation of the N–H···S hydrogen bond, the additional carbonyl oxygen was fixed in its crystallographic position, as reported in [30]. In model C, the backbone of the C116-N117-I118-H119 tetrapeptide was added. The same Cα atoms as in models A and B were kept fixed, as well as the two carbonyl oxygens in order to maintain the two N–H···S interactions. For residues N117 and I118, side chains were replaced by hydrogen atoms in order to save computation time. Each model was optimized in the absence or presence of superoxide anion or hydroperoxide bound to the iron atom. A model was also considered with a hydroxide anion bound to iron, as a model of the $Fe^{3+}$ resting state. Calculations were performed using the Gaussian03 package [31] and geometries were optimized with the B3LYP [32, 33] hybrid density functional using a double-ζ quality basis set (hereafter B1). The 6-31G* basis set was used for all atoms but Fe, N, O and S of the first coordination sphere of the iron. For these N, O and S atoms of the first coordination sphere of the iron and involved in hydrogen bonds, the 6-31+G* basis set was employed. For iron, the lanl2dz with ECP [34] was completed by a set of f polarization functions of exponent 2.462 and a set of d diffuse function of exponent 0.0706 [35]. Energies were computed with B3LYP and the all-electron triple-ζ 6-311+G** (called B2 in the following) basis set on previously optimized geometries. Polarization effects on the environment were accounted for by the mean of the PCM implicit model of solvent ([36] and references therein), with a dielectric constant of 5.75 and a probe radius of 2.7, at the B1 level. The proton affinity of the different models was explored by computing the energetics of the reaction leading to its protonated counterpart. This was made by using a solvation energy of 260 kcal/mol [37]. All systems have been studied in their high-spin state (S=2 for $Fe^{2+}$ resting state, S=5/2 for $Fe^{2+}$-$OO^{-}$, $Fe^{3+}$-OOH and $Fe^{3+}$-OH), consistent with previous results obtained on similar models using comparable methods [19, 20, 30].

## RESULTS

*Electronic absorption spectroscopy*. The UV-visible absorption spectra of the purified *D. baarsii*, I118A, I118D and I118S SOR mutants exhibited the characteristic absorption bands at 370 nm and 503 nm ($\varepsilon_{503nm}$ = 4.4 $mM^{-1}$ $cm^{-1}$) (data not shown), arising from the ferric iron center I [6]. When treated with $K_2IrCl_6$, the spectra of the mutants exhibited an increase in absorbance in the 500-700 nm region, reflecting oxidation of the iron center II [16]. As reported for the wild-type SOR [16], the absorption band maximum of the ferric iron center II of the I118 mutants was blue-shifted at alkaline pHs (Supplemental data Figure S1). This process, called alkaline transition, reflects the displacement of the glutamate 47 ligand on the sixth coordination position of center II ($Fe^{3+}$-E47, maximum of absorption at 644 nm) by a hydroxide ion, to form an $Fe^{3+}$-OH species (maximum of absorption at 560 nm) [16, 39]. As shown in Table 1, when compared to the wild-type SOR, the apparent p$K_a$ associated with this process was significantly higher in the I118S SOR mutant. This effect was less pronounced for the I118A and I118D mutants. In addition, the absorbance band maxima of the $Fe^{3+}$-E47 forms of the I118A, I118D and I118S SOR mutants were shifted by +7 nm (-167 $cm^{-1}$), +4 nm (-96 $cm^{-1}$) and +7 nm (-167 $cm^{-1}$), respectively, when compared to the wild-type SOR (Table 1). Since the SOR proteins were found highly unstable at pH values higher than 10.2, the absorbance band maxima of the $Fe^{3+}$-OH forms of the I118 mutants could not be determined, but we deduced from the spectra recorded at pH 10.2 that they were below 580 nm (Supplemental data Figure S1).

*EPR spectroscopy*. The 4 K EPR spectra of the as-isolated I118 SOR mutants displayed resonances at g = 7.7, 5.7, 4.1 and 1.8, similar to those obtained for the wild-type protein [6] (data not shown). These resonances were typical of those of an $Fe^{3+}$ ion in a distorted tetrahedral $FeS_4$ center and originated from center I [6, 27]. The 4 K EPR spectrum of the I118 SOR mutants oxidized with a slight excess of $K_2IrCl_6$ exhibited an additional signal at g = 4.3, identical to that observed for the oxidized wild-type SOR (data not shown). This g = 4.3 signal was attributed to the oxidized center II, with a rhombic (E/D = 0.33) high-spin (S = 5/2) ferric ion [6, 27]. No other signal was detected (data not shown).

*Redox potentials*. Potentiometric titrations of the I118 SOR mutants were monitored by electronic absorption spectroscopy at pH 7.6. The iron center II of each mutant was oxidized by addition of a stoichiometric amount of $K_2IrCl_6$ and the redox dependence of its absorption spectrum was studied by successive additions of sodium dithionite (Supplemental data Figure S2). The reversibility of the redox process was checked by re-oxidizing the fully reduced protein by addition of $K_2IrCl_6$, and was observed in all instances (data not shown). As shown in Table 1, the I118 A and D mutations had no significant effect on the redox potential of center II at pH 7.6, whereas the I118S mutation slightly increased it.

*FTIR spectroscopy*. FTIR difference spectroscopy was used to evaluate the impact of the mutations on the structural changes accompanying the oxidoreduction of the iron center II. The reduced minus oxidized FTIR difference spectra recorded with the wild-type and the I118S SOR mutant were superimposed in Figure 2a. The spectra are very similar, notably below 1450 $cm^{-1}$, and the IR signatures of the histidine ligands (1109 $cm^{-1}$ for reduced and 1097 $cm^{-1}$ for oxidized SOR) [40] were not significantly altered in the mutant. No IR band could be assigned to the side chain of S118, indicating that the redox state of the iron does not impact the vibrational properties of the S118 side chain. Significant differences with the wild-type were revealed in the 1700-1450 $cm^{-1}$ region for the I118S mutant. These differences are evidenced in the mutant-minus-wild-type difference spectrum recorded in $H_2O$ (Figure 2b). They consist in signals appearing at 1670-1660 (+) / 1683, 1639 (-) $cm^{-1}$ and at 1554, 1504 (+) / 1543, 1523 (-) $cm^{-1}$, that could be confidently assigned to amide I (ν(C=O)) and amide II (ν(CN)+δ(NH)) modes of peptide bonds, since they shifted by -3 to -11 $cm^{-1}$ (ν(C=O)) and by more than -50 $cm^{-1}$ (ν(CN)+δ(NH)), respectively, in spectra recorded in $D_2O$ (Figure 2c). The bands observed between 1680 and 1630 $cm^{-1}$ could result from frequency shifts of the ν(C=O) mode of peptide groups in the mutant, as compared to the wild type. The I118A mutation induced very similar changes in the FTIR difference spectrum, associated with the reduction of the iron center II, notably in the amide II region where the ν(CN)+δ(NH) modes contribute, while only part of these changes was observed for the I118D mutation (Supplemental data, Figure S3). These data show that the I118S and I118A mutations have no major global impact on the structure of the SOR active site, but that they significantly affect the IR mode frequencies of at least one peptide bond around the iron center II.

*Resonance Raman (RR) spectroscopy.* Figure 3A shows the 15 K RR spectra of $K_2IrCl_6$-oxidized wild-type and I118 SOR mutants, excited at 647.1 nm, which allowed a resonance with the $S^-$(Cys) → $Fe^{3+}$ charge transfer (CT) transition (644-651 nm). The RR spectra of SORs excited in such a condition show a number of bands arising from Fe-S stretch coupled with vibrations of the Cys residue [27, 41]. This effect is very similar to that observed for blue copper proteins and Fe-S proteins for which the frequency, number and relative intensity of the observed RR bands are dependent on the Fe-S bond strength and the conformation of the Cys side chain [42-46].

The observed RR frequencies of the wild-type *D. baarsii* and its I118 mutants are listed in Table 2. In the RR spectra of *D. vulgaris* and *D. baarsii*, excited at 647.1 nm, four predominant bands are observed between 200 and 800 $cm^{-1}$. These bands peak at 299, 314, 357 and 743 $cm^{-1}$ for the *D. vulgaris* SOR [27], and at 299, 317, 358 and 742 $cm^{-1}$ for *D. baarsii* one (Figure 3A). On the basis of $^{34}S$ and $^{15}N$ labeling [27], the 299 $cm^{-1}$ line was assigned to a mode involving primarily the Fe-S(Cys) stretch (ν(FeS)). The 314 and 357 $cm^{-1}$ bands were attributed to major contributions of deformations of the $SC_{\beta}C_{\alpha}$(Cys) grouping ($\delta(SC_{\beta}C_{\alpha})$), and of the Cys side chain ($\delta(SC_{\beta}C_{\alpha}$(C(O))N(H)C(O)), respectively. A mode mainly corresponding to the S-$C_{\beta}$ stretch ($\nu(SC_{\beta})$) was associated with the 743 $cm^{-1}$ band [27]. Since the SORs from *D. baarsii* and *Desulfovibrio vulgaris* belong to the same 2Fe-SOR family and exhibit very similar spectroscopic properties [27], we have analyzed our RR data on the basis of the band assignments proposed for the *D. vulgaris* enzyme [27].

The 299 $cm^{-1}$ band of wild-type SOR from *D. baarsii* was significantly shifted to 302 $cm^{-1}$ in the spectra of the I118A, I118D and I118S mutants (Figures 3A and Supplemental data Figure S4; Table 2). These upshifts were also clearly identified in the "wild-type minus I118 mutant" difference spectra (Figure 3B). When the intensity of the mode involving a major contribution of $\nu(SC_{\beta})$ is kept constant (742-744 $cm^{-1}$), one can observe an increase in relative intensity of the 302 $cm^{-1}$ band of the mutants when compared to the 299 $cm^{-1}$ line of the wild-type SOR (Figure 3A). In the RR spectra of blue copper systems and Fe-S proteins excited in S(Cys)-to-metal CT transitions, the most intense peak is generally the one exhibiting the largest $^{34}S$ isotope sensitivity and thus having the predominant ν(Fe-S) character [42-46]. The same observations hold for the red-excited-RR spectra of 1Fe-SORs and 2Fe-SORs [23, 27, 41]. Displacement of atoms along the Fe-S coordinate in the electronic excited state, involving the $S^{-} \rightarrow Fe^{3+}$ CT, is responsible for an Fe-S stretch being the most intense feature in the RR spectra excited at 647.1 nm. The most intense line of the RR spectra of 2Fe-SORs is observed in the 290-310 $cm^{-1}$ region and exhibit the highest $^{34}S$ sensitivity. The relative increased intensities of the 302 $cm^{-1}$ bands of the I118 mutants may be assigned to an increased resonance condition given the redshifts of the CT band (Table 1). Thus, the most intense RR band in the 300 $cm^{-1}$ region is a sensitive indicator of the Fe-S bond strength [23]. The upshifts from 299 $cm^{-1}$ for the wild-type enzyme to 302 $cm^{-1}$ for the three I118 mutants support an increased strength of the axial Fe-S(Cys) bond in the mutants (Table 2 ).

The 314-317 $cm^{-1}$ band of the wild-type SOR originates from kinematic and vibronic coupling of the internal Cys deformations ($\delta(SC_{\beta}C_{\alpha})$) with the Fe-S stretch [27]. Due to the strong decrease in RR activity of the 317 $cm^{-1}$ band, its position in the spectra of the I118 mutants is difficult to establish (Figure 3A). The spectral deconvolutions indicate that its homolog corresponds to either the weak band at 314-316 $cm^{-1}$ bands or the weak or medium bands at 308-309 $cm^{-1}$ (Supplemental data Figure S4). This decreased RR activity is hardly compatible with the redshifts of the $S^{-}$(Cys) → $Fe^{3+}$ CT (Table 1). It would rather correspond to a change in coupling of the $\delta(SC_{\beta}C_{\alpha})$ and $\nu$(Fe-S) mode, i.e. a decreased contribution of the Fe-S stretch in the normal mode composition of the 308-317 $cm^{-1}$ band. In the RR spectra of the blue copper proteins and its model compounds, such a change in mode coupling originates from a modification of the (metal-S-$C_{\alpha}$)-(S-$C_{\beta}$-$C_{\alpha}$) dihedral angle [43]. The RR bands of the 340-410 $cm^{-1}$ region were also significantly affected in the spectra of the I118 mutants (Figure 3 A, B; Supplemental data Figure S4). These deformation modes of the cysteine ligand with the peptide backbone to which it is attached [27, 41] (Table 2) exhibit changes in frequency and/or in intensity, supporting some local protein reorganization. The spectral changes of the 310-410 $cm^{-1}$ RR bands are thus all consistent with a conformational change of the cysteine side chain and the peptide backbone in its proximity.

In the high-frequency region of RR spectra, the 742 $cm^{-1}$ band of the wild-type SOR is gradually upshifted at 743 $cm^{-1}$ for the I118D mutant and at 744 $cm^{-1}$ for the I118A and I118S mutants (Figure 3A). In the difference spectra (Figure 3B), the s-shaped signal in the 710-760 $cm^{-1}$ region illustrates the marked sensitivity of the $\nu(SC_{\beta})$ mode to the I118A and I118S mutations. Its increased frequency reflects an increase of the electron density of the S-$C_\beta$ bond.

Therefore, for the I118 SOR mutants, changes in energy and in intensity of the RR modes involving the Fe-S-$C_\beta H_2$- bonds were observed. These changes are consistent with a conformational change of the Cys side chain inducing a change in H-bonding. On the basis of the frequency upshifts of the primarily ν(Fe-S(Cys)) and ν(S-$C_\beta$) modes, the Fe-S and S-$C_\beta$ bond strengths are increased, indicating a decreased H-bonding state of the S(Cys) site.

Weak bands in the 210-240 $cm^{-1}$ region have been attributed to stretching modes of the Fe-N(His) bonds (ν(FeN(His))) [27, 41]. Although in the I118 SOR mutants, intensity variations were detected for these bands, no significant frequency changes were observed (218-220 and 240-241 $cm^{-1}$) (Figure 3 and Table 2). Thus, the equatorial Fe-$N_4$(His) coordinate is not significantly affected by mutations of the I118 residue.

*Pulse radiolysis study of the I118 SOR mutants.* The reaction of the I118 SOR mutants with $O_2^{\bullet-}$ was studied by pulse radiolysis, which allowed the observation of transient intermediates formed during the reaction course [18, 19]. The kinetics of the reaction were followed spectrophotometrically between 450 and 700 nm and the protein was present in a large excess with respect to $O_2^{\bullet-}$, ensuring pseudo first-order conditions. A 425 nm cut-off filter was positioned on the light beam in order to avoid photochemical reactions involving the reaction intermediates [18]. Similarly to the wild-type protein [16, 18, 19], all the I118 SOR mutants reacted very rapidly with $O_2^{\bullet-}$ to form, 50 μs after the pulse, the first reaction intermediate T1, proposed in the case of the *D. baarsii* enzyme to be an $Fe^{2+}$-superoxo species (Scheme 1, [19]). The absorbance spectrum of T1 was not significantly affected by the I118 mutations, exhibiting a broad band centered at 600-610 nm, with an extinction coefficient of about 2.5 $mM^{-1}$ $cm^{-1}$ (Supplemental data Figure S5). Second-order rate constants for the formation of T1 ($k_1$) were determined to be (1.2 ± 0.2) x $10^9$ $M^{-1}$ $s^{-1}$, (1.1 ± 0.3) x $10^9$ $M^{-1}$ $s^{-1}$ and (1.4 ± 0.3) x $10^9$ $M^{-1}$ $s^{-1}$ for the I118A, I118D, and I118S SOR mutants, respectively. These values are very similar to that determined for the wild-type protein ((1.0 ± 0.2) x $10^9$ $M^{-1}$ $s^{-1}$) [16, 19]. For the three mutants, $k_1$ was found to be pH-independent between pH 5 and pH 9.5 (data not shown), as reported for the wild-type protein [16, 19].

For the wild-type protein, the decay of T1 led to the formation of a SOR $Fe^{3+}$-OH/$H_2O$ species (P1 in Scheme 1), with a rate constant $k_2$. Log $k_2$ was found to decrease linearly with increasing the pH value for pHs < 8.0 and to be pH-independent at pH > 8, with a transition at pH 8. At pH < 7, the I118A, I118D and I118S SOR mutations did not markedly affect the values of $k_2$ and its pH-dependence (Figure 4 and Supplemental data Figure S6). However, for the I118 mutants at pH > 7.0, the pH transition was lowered to 7 and $k_2$ became pH-independent. Thus, the pH transition appeared at a more acidic pH in the mutants when compared to the wild-type SOR (Figure 4 and Supplemental data Figure S6). Consequently, the $k_2$ values at pHs above 8 were increased by a factor of about 4 for the I118 SOR mutants in comparison to that of the wild-type SOR.

The spectra of P1 for the I118 SOR mutants were reconstructed at pH 6.0 and pH 9.5, 10 ms and 180 ms after the beginning of the reaction, respectively (Figure 5 and Supplemental data Figure S7). For the wild-type SOR, this spectrum depends on the pH value, reflecting the acid-base equilibrium between the $Fe^{3+}$-$OH_2$ and $Fe^{3+}$-OH species (maxima of absorbance at 640 and 560 nm, respectively), with an estimated p$K_a$ value of 7.0 [19] (Figure 5 and Supplemental data Figure S7). At pH 6.0, for the I118A, D and S SOR mutants, the P1 spectra were almost superimposable to that of the wild-type SOR at the same pH (data not shown). However, for the I118A, D and S SOR mutants at pH 9.5, these spectra differed from that of the wild-type SOR at the same pH, exhibiting a broad band with a maximum at 600-625 nm. Such a broad absorbance band is consistent with the presence of a mixture of $Fe^{3+}$-Glu (644 nm) and $Fe^{3+}$-OH species (560 nm) at pH 9.5, which was similarly observed for the wild-type protein, but at a lower pH (pH 8.5, [19]). These data suggest that the p$K_a$ value of P1 in the I118 mutants was increased compared to that of the wild-type SOR.

For the wild-type SOR, the evolution of P1 to the final reaction product P2 (rate constant $k_3$) was shown to correspond to the binding of the E47 carboxylate side chain to the ferric iron site, in place of the OH/$OH_2$ ligand (Scheme 1) [19, 38]. The pH dependence of $k_3$ was proposed to reflect a general acid catalysis, associated with the p$K_a$ value of the $Fe^{3+}$-OH/$Fe^{3+}$-$OH_2$ equilibrium (P1), where the aquo ligand was more easily displaced by the carboxylate side chain of E47 than the hydroxo ligand [17, 19]. For the I118A, D and S SOR mutants, the pH-dependence of $k_3$ was still consistent with such a general acid catalysis, with however an apparent p$K_a$ value higher than that determined for the wild-type protein, 9.1 versus 7.0, respectively (Figure 6 and Supplemental data Figure S8). These data suggest that the I118A, D and S SOR mutations induced a stabilisation of the $Fe^{3+}$-$OH_2$ species in the P1 species, in comparison to the wild-type SOR.

The spectrum of the final product P2 was reconstructed at pHs 6.0 and 9.5 for the I118 SOR mutants (Figure 7 and Supplemental data Figure S9). At pH 6.0, the mutants exhibited a spectrum similar to that of the wild-type protein, with an absorbance band centred at 644 nm, associated with an $Fe^{3+}$-Glu species [16] (Figure 7 and Supplemental data Figure S9). At pH 9.5, the spectrum of the final product of the wild-type SOR exhibited an absorbance band peaking at 560 nm, characteristic of an $Fe^{3+}$-OH species [16, 39]. For the I118A, D and S SOR mutants at pH 9.5, the spectrum of the final product, exhibited a broad absorbance band with maximum at 600-625 nm, consistent with the presence of a mixture of $Fe^{3+}$-Glu (644 nm) and $Fe^{3+}$-OH species (560 nm) at this pH. These results suggest an increase of the apparent p$K_a$ value of the final $Fe^{3+}$-E47 species for these three SOR mutants, in comparison to that of the wild-type SOR. This is in agreement with the increase of the apparent p$K_a$ value of the alkaline transition determined for these three mutants (Table 1).

*Pulse radiolysis study of the E114A SOR mutant.* When compared to the previous data obtained for the E114A SOR mutant [23], the I118A, D and S mutations appeared to have opposite effects both on the strength of the Fe-S bond and on the protonation rate of the T1 intermediate, as observed by pulse radiolysis. However in reference [23], the pulse radiolysis experiments were carried out in conditions where photochemical reactions likely occurred at the level of the reaction intermediates [18]. Here, the reaction kinetics of the E114A mutant with $O_2^{\bullet -}$ were reinvestigated in the presence of a 425 nm cut-off filter in order to avoid any photochemical processes [18]. Similarly to the wild-type and the I118 SOR mutants, the reaction of the E114A mutant with $O_2^{\bullet -}$ led to the formation of two transients T1 and P1, before evolving to the final reaction product P2 (data not shown). The rate constant of T1 formation ($k_1$) was not affected by the E114A mutation and was found to be pH-independent between pH 5.5 to 9.5 (data not shown). At pHs between 5.5 and 8.5, the decay of T1 (rate constant $k_2$) was found to be directly proportional to the $H^+$ concentration (Supplemental data Figure S10.A). However in this pH range, the $k_2$ values determined for the E114A mutant were found to be 6-9 times smaller than those determined for the wild-type protein. At pH > 8.5, for the E114A mutant, $k_2$ slightly increased with decreasing $H^+$ concentration, with values from 7 to 10 times smaller than those found for the wild-type SOR (Supplemental data Figure S10.A). These data show that the E114A SOR mutation decreases the rate of protonation of T1, as it was previously reported in [23]. This indicates that the photochemical processes occurring in this former study did not affect the rate of $k_2$. As shown in Supplemental data Figure S10.B, the pH-dependence of $k_3$ for the E114A SOR mutant exhibited an apparent $pK_a$ value very similar to that of the wild-type SOR. However, the E114A mutation induced a decrease of the $k_3$ values by a factor of 2.5 to 10 between pH 5.5 and pH 9.6 (Supplemental data Figure S10.B).

*DFT Calculations.* In order to rationalize the effects of H-bonds toward the sulfur ligand on the reactivity of SOR with $O_2^{\bullet-}$, three models corresponding to the five-membered $Fe^{2+}$ form, the T1 intermediate ($Fe^{2+}$-$OO^{\bullet}$) and its subsequently protonated species ($Fe^{3+}$-OOH), have been studied by DFT calculations. A fourth structure, $Fe^{3+}$-OH, has also been considered to account for a model of the ferric state of SOR after formation of hydrogen peroxide. Each intermediate has been modeled by three models, each with a different number of N-H…S interactions. Model A has no H-bond on the sulfur ligand, model B has one H-bond, and model C has two H-bonds (wild-type SOR). Figure 8 shows the structure of the three models A, B and C for the $Fe^{2+}$-$OO^{\bullet}$ species (T1). Discussed geometric parameters (Fe-S and Fe-O distances) as well as charges and spin population of Fe and S atoms are presented in Table 3 for the four species $Fe^{2+}$, $Fe^{2+}$-$OO^{\bullet}$, $Fe^{3+}$-OOH and $Fe^{3+}$-OH, each described by models A, B and C. A more complete set of data is available as Supplemental data, Tables S1, S2 and S3.

For the five-membered $Fe^{2+}$ structure, models A, B and C showed good agreement with the corresponding PDB structure of SOR (2IJ1) [14]. In term of geometry, the Fe-S bond is slightly affected by the number of H-bonds toward S with a distance of 2.35 Å in A, 2.37 Å in B and 2.39 Å in C (2.41 Å in PDB). Consistently, a slight increase of the atomic charge and spin population of Fe, accompanied by a decrease of respective values for S, is found when including one and then two H-bonds toward S. These trends are significant and indicate that the Fe-S bond is weakened by H-bonds toward S. Similar observation were made for the $Fe^{2+}$-$OO^{\bullet}$ and $Fe^{3+}$-OOH species.

For the $Fe^{2+}$-$OO^{\bullet}$ species (models A, B, C), the -$OO^{\bullet-}$ moiety bears a total charge of ca. -0.8 (i.e. the substrate keeps most of its negative charge) and has a spin population slightly larger than 1.0. Upon binding of the -$OO^{\bullet-}$ moiety to the ferrous iron, the spin population as well as the atomic charge do not vary significantly. Consistently with previous findings [19], the iron atom keeps its formal oxidation state of +2, while superoxide remains a radical anion.

Nevertheless, a sizeable increase of the Fe-S distance occurs when $OO^{\bullet -}$ binds to Fe (i.e.: passing from the five-membered $Fe^{2+}$ to the $Fe^{2+}$-$OO^{\bullet}$ one): 0.20 Å in model A, 0.24 Å in B and 0.32 Å in C. This decrease in Fe-S interaction can be related to a trans influence between S and superoxide.

As for the five-coordinate $Fe^{2+}$ species, comparison of structures for models A, B and C shows that the Fe-S bond is lengthened if the number of H-bonds is increased. This trend is noticeable, since the Fe-S distance is 2.55 Å in A, 2.61 Å in B and 2.72 Å in C. Consistently, the Fe-O distance decreases with the increase of the Fe-S one: the values for the Fe-O bond length are 2.29 Å in model A, 2.23 Å in B and 2.20 Å in C. The presence of H-bonds thus modulates the trans influence between the sulfur ligand and the superoxide.

Protonation of the distal oxygen of the $Fe^{2+}$-$OO^{\bullet}$ species leads to the iron-hydroperoxo intermediate ($Fe^{3+}$-OOH), where the hydroperoxide moiety has an overall charge of -1.1, with a spin population of about 0.3 in the three models A, B and C. The substrate protonation is accompanied by an increase of the atomic charge of the iron in the three models (e.g.: in model C, the charge of iron is 1.34 for the $Fe^{2+}$-$OO^{\bullet}$ species and 1.52 in the $Fe^{3+}$-OOH one). Accordingly (see Table 3), the sulfur ligand becomes less negative in the three models after protonation. The spin populations of both the Fe and S atoms also increase during this process. This shows that the substrate moiety loses its radical character by abstracting one electron after protonation, thus leading to a formal $Fe^{3+}$ covalently bound to an $OOH^{-}$ moiety. In term of geometry, the Fe-S and Fe-O bonds are shorter than in the $Fe^{2+}$-$OO^{\bullet}$ structure. This is consistent with an iron atom formally losing one electron.

When comparing the models A, B and C for the $Fe^{3+}$-OOH species, the general trends observed in the previous $Fe^{2+}$-$OO^{\bullet}$ intermediates are found again. The presence of H-bonds toward sulfur makes the Fe-S bond longer and the Fe-O one shorter. However, this effect remains much more limited than for the $Fe^{2+}$-$OO^{\bullet}$ species. For the $Fe^{3+}$-OOH species, the changes in Fe-S and Fe-O bond lengths are found in a range of 0.04-005 Å in the three models. Interestingly, while H-bonds have a limited effect on geometries, their influence on the electronic structure is more important. Hence the spin population of sulfur is larger in model A than in models B and C with respective values of 0.49, 0.38 and 0.33, respectively. Consistently, the charge separation between sulfur and iron is enhanced by the presence of H-bonds (sulfur becomes more negative and iron more positive (see charges, Table 3). Again the presence of H-bonds weakens the Fe-S interaction, which in turn strengthens the Fe-O one. Indeed, the charge of the oxygen atom bound to iron decreases while its spin population increases, going from A to B and C. As can be seen in Table 3**,** the last structure studied, the $Fe^{3+}$-OH species which is formed alter the release of $H_2O_2$, shows similar trends as already pointed. The Fe-S bond is shortened while the Fe-O one becomes longer when adding H-bonds toward S. The effect of H-bonding is of comparable magnitude as previously noted with a variation of +0.08 Å for Fe-S between A and -0.05 Å for Fe-O between models A and C. Charges and spin population are also in agreement with a reinforced Fe-S interaction when H-bonds are absent. These observations are in agreement with the increased strength of the Fe-S bond in the I118 SOR mutants observed by RR spectroscopy (Figure 3). RR also showed an increase of the $C_\beta$ - S interaction. However, due to constraints applied to our models (fixed $C_\alpha$), our calculations did not provide any sizeable modification of the $C_\beta$ - S distance, which remains at values close to 1.86 Å in the three models A, B and C of the $Fe^{3+}$-OH resting state.

Finally, we also have studied the energetics for the protonation process of the $Fe^{2+}$-$OO^{\bullet}$ intermediate to form the $Fe^{3+}$-OOH species, which has been computed for each model A, B and C. For the models A, B and C, the protonation process is found exothermic by 41.2, 36.2 and 32.6 kcal/mol respectively when using -260 kcal/mol as a reference for the solvated proton [37] and B3LYP/B2//B3LYP/B1 plus solvation effects for calculated values. These values are consistent with the -36 kcal/mol reported by Dey *et al.* for the same reaction, using a similar approach [30]. Comparison of the values found for the three models shows that H-bonds toward the S atom make the protonation process less exothermic. Compared to model C (wild-type SOR), the reaction is more exothermic in model B (I118 SOR mutants) by 5.6 kcal/mol and by 8.9 kcal/mol in model A (no H-bonds). Hence, thermodynamically, the presence of H-bonds disfavors the protonation process of the first reaction intermediate T1. These data are in agreement with the experimental observations presented in Figure 4, showing that the I118 SOR mutations, which impaired one H-bond on the sulfur ligand, induced an increase of the protonation rate of the T1 intermediate.

## DISCUSSION

The presence of H-bonds on the sulfur atom of cysteine ligands is prominent in iron proteins, e.g. in heme-thiolate proteins like cytochrome P450 or [Fe-S] cluster proteins [24-26]. In SORs, two peptide NH groups from L/I118 and H119 are within H-bonding distance of the C116 axial sulfur ligand (Figure 1, [11-14]), and they have been proposed to play an important function in SOR [27, 30]. However, up to now, no experimental data were reported concerning their role in catalysis. Site directed mutagenesis is a suitable technique for such investigations, however some limitations have to be taken into account. As highlighted in the Introduction section, the imidazole side chain of H119 is a ligand of the iron and consequently this position cannot be mutated without introducing large perturbations, and possibly demetallation of the active site. For the buried I118 position, mutation into a proline residue to probe the function of its peptide NH group could also induce large conformation changes in the protein. In fact, in the different amino-acid sequences of SOR available to date, no proline residue was found in this position. Here we have mutated the I118 residue into A, D, and S, assuming that the modifications of the side chain of I118 could impact the orientation and the strength of the peptide NH hydrogen bond to the sulfur ligand.

Such specific effects of the I118 mutations on the peptide NH H-bond were supported by FTIR spectroscopy, a very sensitive technique to investigate the structural modifications accompanying the change in the redox state of the SOR iron site [22, 23]. FTIR showed that whereas the I118 mutations had no major impact on the active site structure, they significantly affected the conformation of at least one peptide bond around the iron active site, which are compatible with a change in hydrogen bond involving peptide NH group(s).

RR spectroscopy further showed that the I118 A, D and S mutations had very specific effects on the cysteine ligand. For these three mutants, the frequencies of the stretching modes involving the Fe-S-$C_\beta H_2$- grouping clearly indicated an increase strength of the $Fe^{3+}$-S(Cys) and S-$C_\beta$ bonds. These data are in agreement with the alteration of at least one of the H-bonds engaged by the S(Cys) ligand, inducing an increase of the electron density on the S(Cys) atom in the I118 mutants. Interestingly, all these small structural alterations can be directly related to the variation in the position of the S $\rightarrow$ $Fe^{3+}$ charge transfer band [27, 47], which was observed in the I118A, D and S mutants (+7, +4 and +7 nm, respectively, Table 1). Since the maximum of the charge transfer band corresponds to the donor-acceptor transition energy of the Fe-S bond, a decreased transition energy implies a more favorable transfer, i.e. that the S(Cys) atom becomes less electronegative via weakening of its H-bonding state, in agreement with our data. These changes in Fe-S bond strength do not however affect the EPR spectrum of the SOR active site, which remains high-spin (S = 5/2) with a rhombic geometry (E/D = 0.33), as reported for the wild-type SOR from *D. baarsii* [6]. No axial feature, such as those reported for the SOR from *P. furiosus* [47] and *T. pallidum* [48] was found in the EPR spectra of the I118 SOR mutants.

Taken together, these data demonstrated that the A, S and D mutations have a similar effect on the SOR active site. They all induced an apparently similar decrease in strength of the H-bond between the peptide NH and the sulfur ligand, which in turn lead to a strengthening of the $Fe^{3+}$-S(Cys) bond. This effect most likely originates from a decrease in size of the side chain of residue 118 that could slightly remodel the structure of the C116-N117-I118-H119 tetrapeptide. This decrease in size has to be significant since we found no frequency change for the modes involving the Fe-S-$C_\beta$(Cys) bonds when we compared the RR spectrum of the wild-type SOR with that of the I118V mutant (data not shown).

Surprisingly, the strengthening of the $Fe^{3+}$-S(Cys) bond observed in the I118A, D and S mutants has almost no effect on the redox potential of the iron active site. Only a small positive variation for the S mutant was observed (Table 1). It should be noted that the redox process occurring at the SOR active site is rather complex, involving both electron and proton transfer and coordination of the carboxylate side chain of E47 to the ferric iron. Dey and coll. [30] suggested by DFT calculations that in SOR H-bonds on the thiolate ligand could have an opposite effect on the oxidation of the ferrous iron and on the subsequent coordination of the E47 ligand to the ferric iron. In addition, the redox potential also depends on the stabilization and/or destabilization of the oxidized and/or reduced form of the redox center and the I118 mutants could also strengthen the $Fe^{2+}$-S(Cys) (ferrous) bond. However, no information on the effect of the I118 SOR mutants on the $Fe^{2+}$-S(Cys) bond is available yet. The ferrous form does not present visible absorption band and could not be investigated by RR spectroscopy.

*The I118 peptide NH hydrogen bond to the sulfur ligand controls the p$K_a$ of the reaction intermediates*

The studies of the reactivity of SOR with $O_2^{\bullet -}$ by pulse radiolysis further showed that the presence of the I118 peptide H-bond on the sulfur ligand has a direct control on the p$K_a$ value of the different intermediates species that are formed trans to the Fe-S(Cys) bond during catalysis. Whereas the initial fast binding of $O_2^{\bullet -}$ to the ferrous iron site to form T1 ($k_1$), was not significantly modified by the I118 A, D and S mutations, they all induced an increase in the rate of protonation of T1 ($Fe^{2+}$-$OO^{\bullet}$ species) at pH > 8.0, where $H_2O$ was proposed to be the proton donor [19]. Since I118 is located on the opposite site of the iron atom compared to the $O_2^{\bullet -}$ binding site (Figures 1 and 8), and since no major structural changes were observed in the mutants by FTIR, the increase in the rate of T1 protonation could be hardly explained by any structural modifications that might facilitate proton transfer from the water molecule to the superoxo moiety of T1. Rather, the faster rate of protonation of T1 could be better associated with a specific increase of its p$K_a$ value. This is in agreement with the Brönsted catalysis equation, which predicts that among a series of bases, the logarithm of the rate constant of protonation of a species is directly proportional to its p$K_a$ value. Such a modification of the p$K_a$ value of T1 could result, through a trans electronic effect, from the strengthening of the Fe-S(Cys) bond in the I118 mutants.

It should be noted that at pH < 8, where $H_3O^+$ was shown to be the proton donor [19], the I118 mutations did not affect the rate of protonation step of T1. This could be in line with the fact that in the presence of a strong acid, the dependence of the rate of protonation of a weak base (T1) to its p$K_a$ value might not be valid any more.

Such an effect of the I118 peptide NH hydrogen bond to the sulfur ligand on the p$K_a$ value of the T1 intermediate was further supported by DFT calculations (Figure 8, Table 3). The calculations showed that for the $Fe^{2+}$-$OO^{\bullet}$ species (T1), the Fe-S(Cys) distance increased significantly in the presence of H-bond(s) around the sulfur ligand, while the Fe-O distance decreased slightly. The presence of H-bonds rendered the protonation process of the $Fe^{2+}OO^{\bullet}$ species less exothermic, by 5.6 kcal/mol and 8.9 kcal/mol in the presence of one and two H-bonds, respectively. The Bell-Evans-Polanyi principle [49, 50] stipulates that for similar reactions, the most exothermic one will have the lowest activation energy. Thus, one can anticipate a lowering of the activation energy of protonation of the $Fe^{2+}OO^{\bullet}$ species in the presence of only one of the two hydrogen bonds, as it is the case for the I118 SOR mutants. These computational data are in full agreement with the experimental observations, which showed that the reaction rate $k_2$ for the protonation of the first reaction intermediate was increased in the I118 mutants compared to wild-type SOR. These data also suggest that the second H-bond to the sulfur ligand coming from the H119 NH amide group should have a similar effect on $k_2$. Thus, the two H-bonds from I118 and H119 to the sulfur ligand should have an additive impact on the p$K_a$ of T1.

Interestingly, a similar effect of the presence of H-bond to the sulfur ligand was also observed by pulse radiolysis on the transient P1, formed following the release of $H_2O_2$ from the active site and associated with $Fe^{3+}$-OH/$Fe^{3+}$-$OH_2$ species in acid-base equilibrium [19]. The data showed that the I118A, D and S mutations induced an increase of the apparent p$K_a$ value of P1 by two pH units (from 7.0 to 9.1, Figures 6 and S8), suggesting that removal or weakening of an H-bond to the sulfur ligand in the I118 mutants induces a stabilisation of the $Fe^{3+}$-$OH_2$ species compared to the $Fe^{3+}$-OH one. Again, and similarly to what was described for the $Fe^{2+}$-$OO^{\bullet}$ species, this supports a trans electronic effect of the sulfur ligand on the p$K_a$ of the $Fe^{3+}$-OH/$Fe^{3+}$-$OH_2$ species, which is modulated by the H-bond on this ligand.

Hence, both theoretical and experimental observations support the hypothesis that the presence of H-bonds toward the sulfur ligand of the active site of SOR disfavors the protonation of both the superoxide adduct of the $Fe^{2+}$-O-$O^{\bullet}$ intermediate, and the OH adduct of the $Fe^{3+}$-OH transient, by decreasing their p$K_a$ value.

*SOR active site finely tunes the strength of the Fe-S(Cys) bond in SOR by opposite effects*

In a previous study, the E114A mutation, which cancels a dipolar interaction between the carboxylic side chain of this buried residue with the Fe-S(Cys) bond, was shown to weaken the Fe-S(Cys) bond [23]. In addition, as previously reported in [23] and confirmed here using conditions where no photochemical processes occurred during the pulse radiolysis experiments [18, 19], the E114A mutation was shown to induce a decrease of the p$K_a$ value of the T1 intermediate, which was directly associated with the weakening of the Fe-S(Cys) bond. Thus, the E114 and I118 mutants have opposite effects of the p$K_a$ of the T1 intermediate, which can be directly associated to their opposite effect on the strength of the Fe-S(Cys) bond.

Altogether, these results demonstrated that in SOR, the p$K_a$ of the first reaction intermediate T1 is finely tuned through a tight control of the strength of Fe-S(Cys) bond. This fine tuning is achieved by a combination of opposite effects on the strength of Fe-S(Cys) bond, resulting at least from two different types of interactions: a dipolar interaction with the E114 carboxylic side chain, and a H-bond between the I118 NH amide group and the sulfur ligand.

Such a tight control of the rate of protonation of T1 to form the $Fe^{3+}$-OOH species is rather surprising at first glance. In fact, for efficient catalysis, one could expect a fastest protonation rate as possible. This might suggest that a too fast formation of the $Fe^{3+-}$OOH species could be somehow detrimental for the SOR activity. Our recent studies underlined that the control of the evolution of the $Fe^{3+}$-OOH intermediate is one of the key features of the SOR activity [14, 19]. In particular, K48, which allows the specific protonation of the proximal oxygen of the $Fe^{3+}$-OOH intermediate, was shown to play an essential role in the formation of the reaction product $H_2O_2$. Otherwise, in the absence of such a specific protonation process, the O-O bond of the $Fe^{3+}$-OOH intermediate is cleaved to form a high valent iron-oxo species [19]. It is thus conceivable that the control of the rate of protonation of T1 by the second coordination sphere residues could favour the correct positioning of K48 toward the proximal oxygen of the $Fe^{3+-}$OOH species. Such hypothesis is supported by the fact that K48 can adopt different conformations around the $Fe^{3+}$-OOH species, as observed in the different X-ray structures of this intermediate trapped in a SOR crystal [14]. Any impairment of this specific protonation process due to a bad positioning of the K48 side chain might promote the formation of an iron -oxo species, as supported by our studies on the K48I SOR mutant [19]. The ability of the I118 SOR mutants to favour the formation of an iron-oxo species in their active sites is currently under investigation.

ACKNOWLEDGMENTS

VN and CB acknowledge support from the Agence Nationale de la Recherche, programme Physique et Chimie du Vivant 2008. This work has been partially supported by the Labex ARCANE (ANR-11-LABX-0003-01).

Supporting Information paragraph, Tables S1-S3, Figures S1-S10.

**Table 1.** Effects of the I118 SOR mutations on the physico-chemical properties of the SOR active site

| SOR protein | Maximum (nm) of the absorption band of the ferric iron center II [1,2] | Apparent p$K_a$ of the $Fe^{3+}$-E47/$Fe^{3+}$-OH transition | Redox potential of the iron center II (mV, vs. NHE) |
|---|---|---|---|
| WT | 644 (1.90) | 9.0 ± 0.1 | 308 ± 11 |
| I118A | 651 (1.90) | 9.3 ± 0.1 | 293 ± 7 |
| I118S | 651 (1.92) | 9.6 ± 0.1 | 334 ± 12 |
| I118D | 648 (1.92) | 9.2 ± 0.1 | 292 ± 7 |

[1] Observed at pHs lower than 8 and corresponding to the $Fe^{3+}$-E47 species [16]

[2] In parentheses, values of the extinction coefficient ($mM^{-1}$ $cm^{-1}$)

**Table 2**. Observed frequencies ($cm^{-1}$) of the RR bands of $K_2IrCl_6$ oxidized *D. baarsii* SOR and its I118 mutants, excited at 647.1 nm. The band assignments are from Clay et al. [27]

| WT | I118S | I118A | I118D | Assignment |
|---|---|---|---|---|
| 218 | 219 | 218 | 220 | ν(FeN(His)) |
| 241 | 241 | 240 | 240 | ν(FeN(His)) |
| 277 | 279 | 280 | 279 | |
| 287 | 289 | 287 | 288 | |
| 299 | 302 | 302 | 302 | ν(FeS) + δ(Cys) |
| 306 | 308 | 309 | 308 | $\delta(SC_\beta C_\alpha)$ + ν(FeS) |
| 317 | 314 | 318 | 316 | |
| 324 | 325 | 325 | 324 | |
| 344 | 349 | 346 | 345 | |
| 358 | 359 | 356 | 356 | $\delta(C_\beta C_\alpha C(O))$ + $\delta(C_\beta C_\alpha N)$ + $\delta(C(O)C_\alpha N)$ + $\delta(C_\alpha NC(O))$ + $\delta(SC_\beta C_\alpha)$ + ν(FeS |
| 381 | 380 | 377 | 380 | |
| 396 | 396 | 395 | 396 | |
| 402 | 404 | 406 | 403 | |
| 435 | 439 | 440 | 438 | |
| 455 | 457 | 454 | 454 | |
| 464 | 463 | 463 | 465 | |
| | 468 | 470 | 469 | |
| 507 | 512 | | 512 | |
| 602 | 599 | 601 | 596 | |
| 613 | 612 | | 611 | |
| 649 | 650 | 647 | 648 | δ(His) |
| 658 | 659 | 658 | 658 | $\nu(C_\alpha N)$ |
| 665 | 665 | 665 | 665 | |
| 703 | 703 | 702 | 701 | |
| 713 | 714 | 713 | 712 | |
| 742 | 744 | 744 | 743 | $\nu(SC_\beta)$ |
| 765 | 764 | 764 | 763 | |
| 777 | 777 | 778 | 777 | |
| 793 | 795 | 793 | 793 | |

**Table 3**. Results of DFT calculations with relevant distances (Fe-S and Fe-O), charges (q) and spin populations (S) of Fe and S atoms. Values are given for the four intermediates studied with no (model A), 1 (model B) and 2 (model C) H-bonds toward sulfur. Geometries are obtained at the B3LYP/B1 level of theory while charges and spin populations are calculated at the B3LPY/B2//B3LYP/B1 level of theory

| | $Fe^{2+}$ | | | $Fe^{2+}$-$OO^{\bullet}$ | | | $Fe^{3+}$-OOH | | | $Fe^{3+}$-OH | | |
|---|---|---|---|---|---|---|---|---|---|---|---|---|
| | A | B | C | A | B | C | A | B | C | A | B | C |
| Fe-S (Å) | 2.35 | 2.37 | 2.39 | 2.55 | 2.61 | 2.71 | 2.47 | 2.49 | 2.52 | 2.50 | 2.54 | 2.58 |
| Fe-$O_p$(Å) | - | - | - | 2.29 | 2.23 | 2.20 | 2.03 | 2.00 | 1.99 | 1.96 | 1.94 | 1.93 |
| $q_{Fe}$ | 1.26 | 1.29 | 1.32 | 1.30 | 1.32 | 1.34 | 1.46 | 1.50 | 1.52 | 1.51 | 1.56 | 1.58 |
| $S_{Fe}$ | 3.49 | 3.51 | 3.53 | 3.51 | 3.52 | 3.53 | 3.69 | 3.72 | 3.74 | 3.74 | 3.77 | 3.79 |
| $q_S$ | -0.49 | -0.50 | -0.54 | -0.58 | -0.57 | -0.60 | -0.29 | -0.37 | -0.42 | -0.34 | -0.41 | -0.46 |
| $S_s$ | 0.23 | 0.20 | 0.18 | 0.13 | 0.11 | 0.08 | 0.49 | 0.38 | 0.33 | 0.42 | 0.32 | 0.28 |

## FIGURE LEGENDS

**SCHEME 1**. Mechanism for the reaction of the SOR from *D. baarsii* with $O_2^{\bullet-}$, as proposed in [19]. The first reaction intermediate, T1, presents a broad absorption band centred at 600 nm ($\varepsilon$ = 2.5 $M^{-1}$ $cm^{-1}$) and was proposed to be a ferrous iron-superoxo species ($Fe^{2+}$-$OO^{\bullet}$) [19]. The second reaction intermediate, T1', a ferric iron-hydroperoxo species, was not experimentally observed, but was deduced from studies on the reactivity of the K48I SOR mutant with $O_2^{\bullet-}$ [19] and from the crystal structures of iron-hydroperoxide species trapped in the active site of SOR [14]

**FIGURE 1**. Active site (center II) of the wild-type SOR from *Desulfoarculus baarsii* from its crystal structure at 1.7 Å resolution [14]. Hydrogen bonds are shown in blue dotted lines between the amide NH main groups of I118 (2.45 Å) and H119 (2.24 Å) and the sulfur atom of the C116 ligand. For clarity, the fourth histidine ligand, H75, is not shown

**FIGURE 2**. Reduced *minus* oxidized FTIR difference spectra of the SOR active site of wild-type (black line) and I118S SOR mutant (red line) from *D. baarsii* (a). Potentials of 650 mV and 250 mV (versus NHE) were applied at the working electrode to obtain the oxidized and reduced states, respectively. The spectra consist of an average of data recorded on 30 electrochemical cycles, 300 scans per cycle, 4 $cm^{-1}$ resolution. Mutant minus wild-type difference spectra calculated from the I118S SOR mutant in $H_2O$ (b) or in $D_2O$ (c)

**FIGURE 3**. A. 200-800 $cm^{-1}$ regions of 15 K RR spectra of the $K_2IrCl_6$ oxidized wild-type and I118 SOR mutants, excited at 647.1 nm. For each spectrum, the contributions of the ice and the laser line at 669 $cm^{-1}$ were subtracted. The presentation of the RR spectra was made using the 742-744 $cm^{-1}$ band as an intensity standard. B. Difference spectra (intensity multiplied by a factor of 2 with respect to the RR spectra in A)

**FIGURE 4**. pH-dependence of the rate constant $k_2$ for the reaction of SORs from *D. baarsii* (100 µM) with $O_2^{\bullet-}$ (6 µM), generated by pulse radiolysis in the presence of 10 mM formate and 2 mM buffer. Xenon lamp with a 425 nm cut-off filter. (O) wild-type SOR; (■) I118S SOR mutant

**FIGURE 5**. Transient absorption spectra (tungsten lamp with a 425 nm cut-off filter) of the second reaction transient P1 at pH 6.0 or pH 9.5, formed at 10 and 180 ms, respectively, after the beginning of the reaction of the SOR from *D. baarsii* (100 µM in 10 mM formate, 2 mM buffer) with $O_2^{\bullet-}$ (9 µM), generated by pulse radiolysis. (O) wild-type SOR, pH 6.0; (□) wild-type SOR, pH 9.5; (●) I118S SOR mutant, pH 6.0; (■) I118S SOR mutant, pH 9.5

**FIGURE 6**. pH-dependence of the rate constant $k_3$ for the reaction of SOR from *D. baarsii* (100 µM) with $O_2^{\bullet -}$ (6 µM), generated by pulse radiolysis in the presence of 10 mM formate and 2 mM buffer. Tungsten lamp with a 425 nm cut-off filter. (O) wild-type SOR; (■) I118S SOR mutant. For both proteins, the pH dependence of $k_3$ reflects a general acid catalysis process, where the p$K_a$ value corresponds to the break down of the curve log $k_3$ versus pH (p$K_a$ value of 7.0 and 9.1 for the wild-type and for the I118S mutant, respectively)

**FIGURE 7**. Absorption spectra (tungsten lamp with a 425 nm cut-off filter) of the final reaction product P2 at pH 6.0 or pH 9.5, formed at 100 ms and 2 s, respectively, after the beginning of the reaction of the SOR from *D. baarsii* (100 µM in 10 mM formate, 2 mM buffer) with $O_2^{\bullet -}$ (9 µM), generated by pulse radiolysis. (O) wild-type SOR, pH 6.0; (□) wild-type SOR, pH 9.5; (●) I118S SOR mutant, pH 6.0; (■) I118S SOR mutant, pH 9.5

**FIGURE 8**. Optimized structures of models A, B and C of the active site of SOR, binding superoxide anion ($Fe^{2+}$-$OO^{\bullet}$ species). Model C includes the two H-bonds toward the sulfur ligand, as found in wild-type SOR. Model B has only one H-bond and model A has no H-bond. Relevant atoms are indicated in model C. H-bonds considered are represented by green dashed lines. Circled atoms were kept fixed during geometry optimization, carried at the B3LYP/B1 level of theory

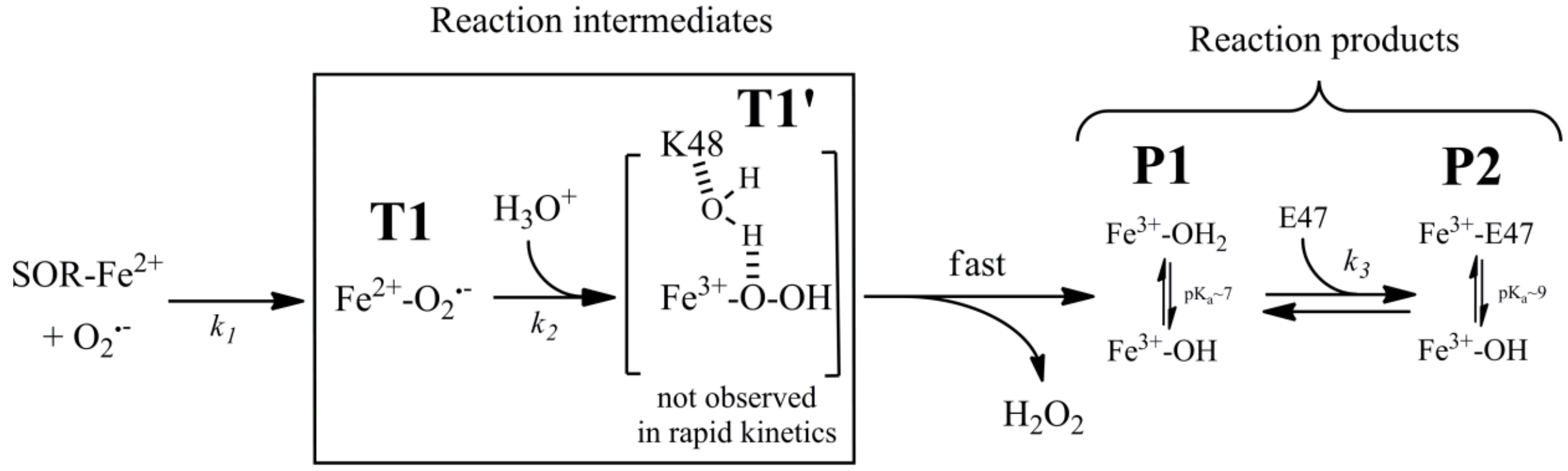

Scheme 1

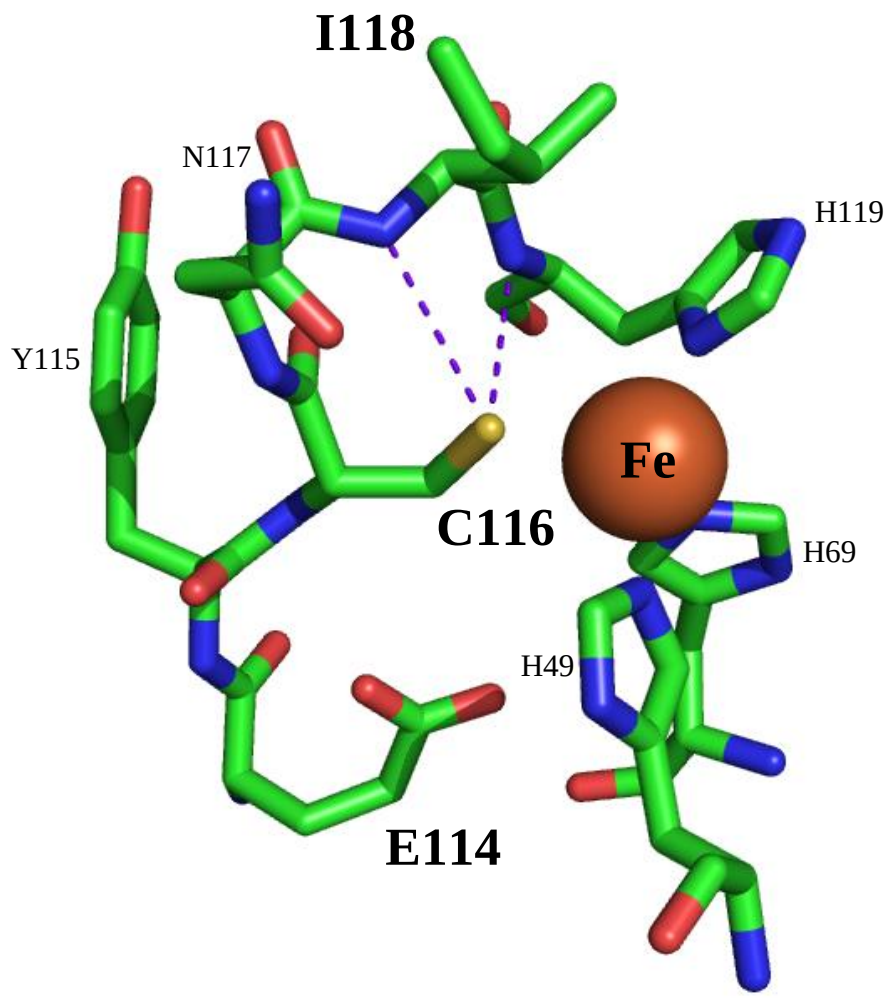

Figure 1

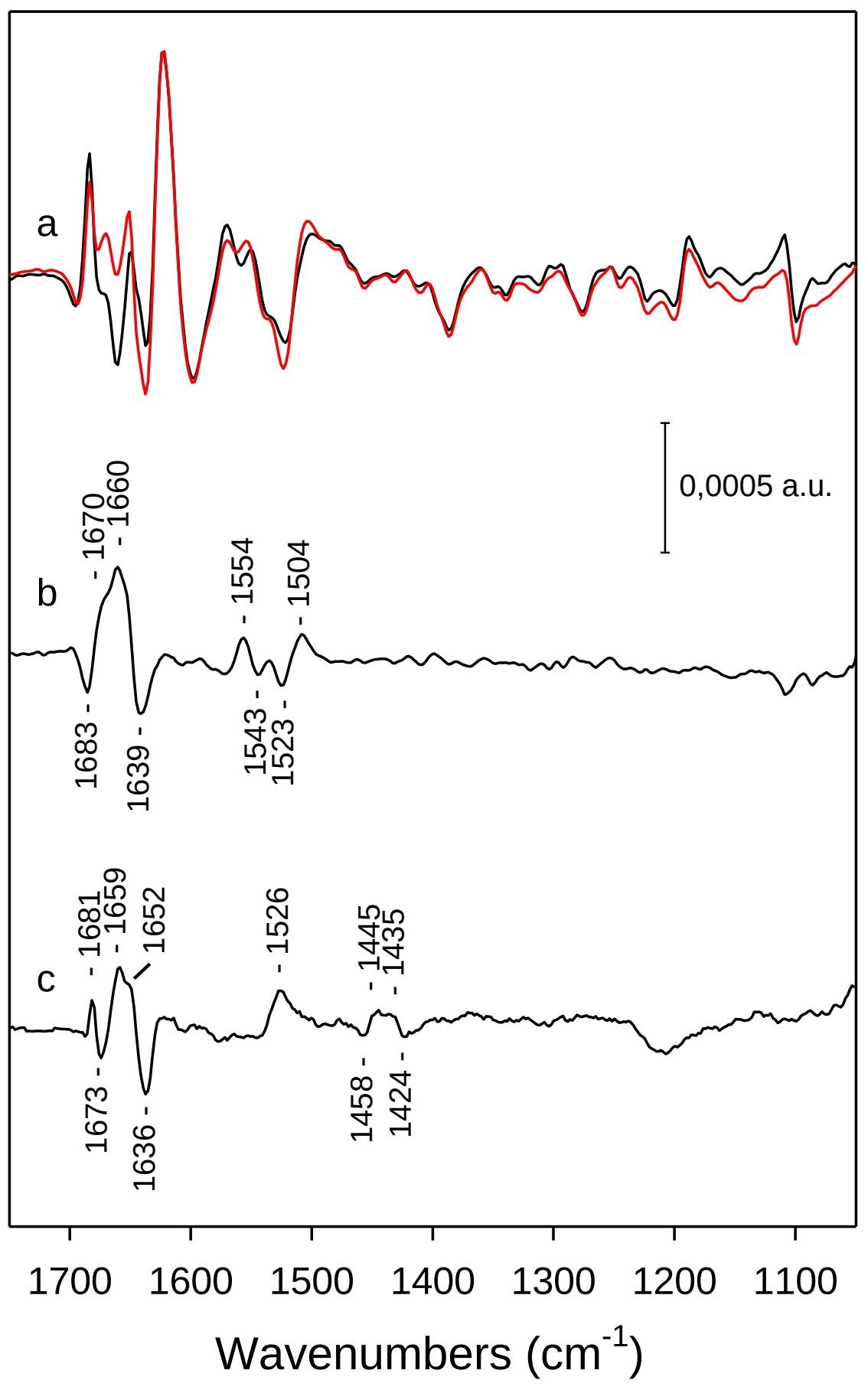

Figure 2

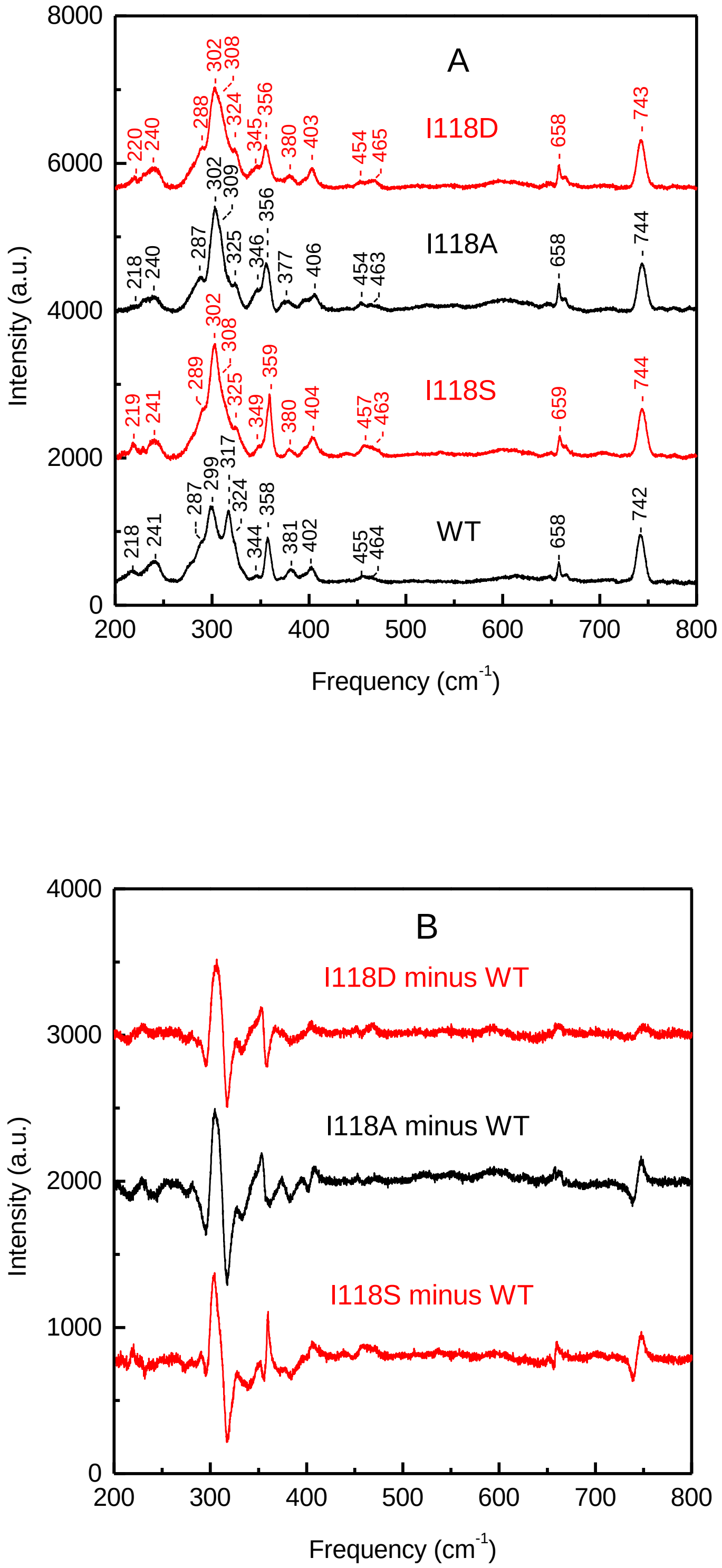

Figure 3

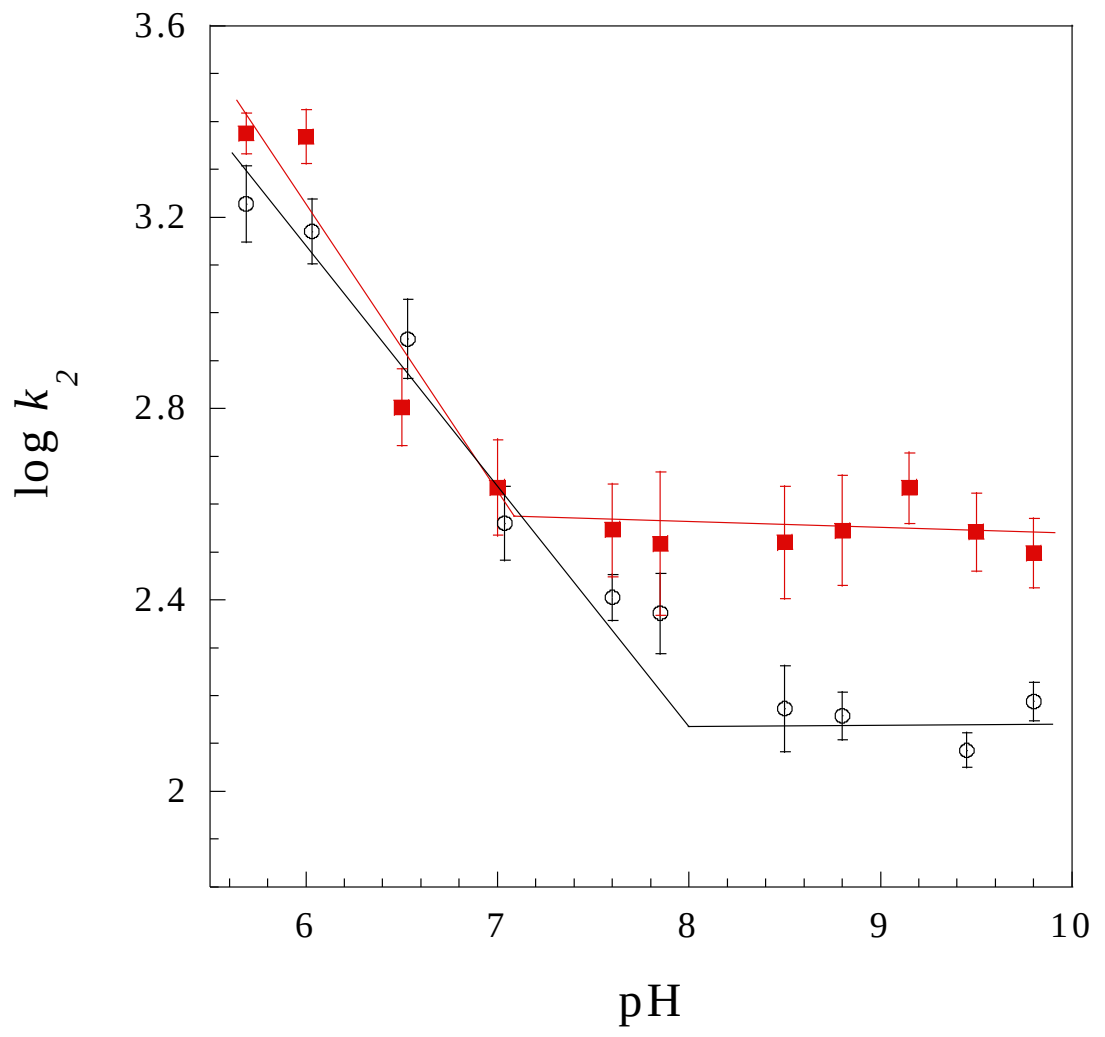

Figure 4

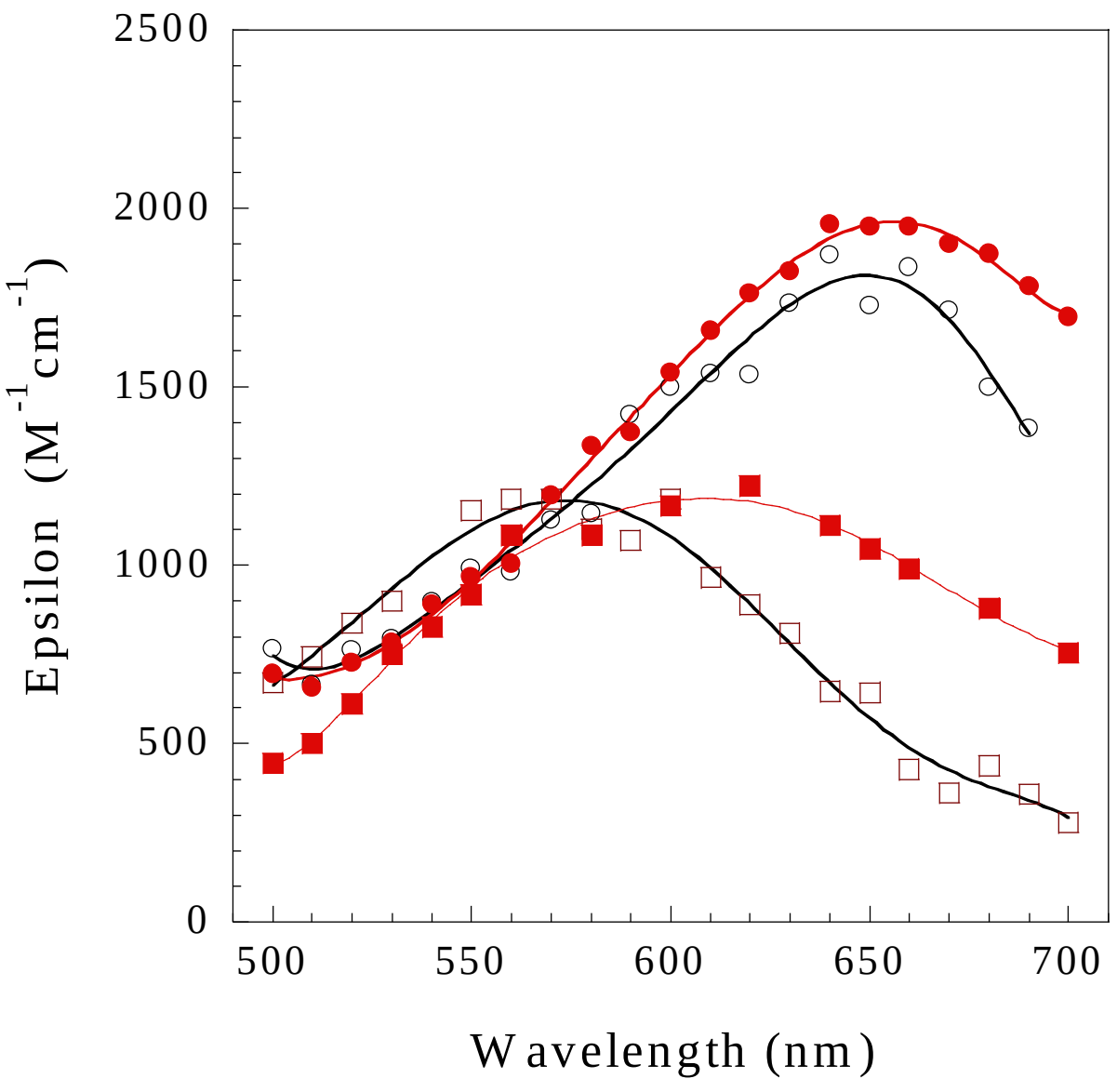

Figure 5

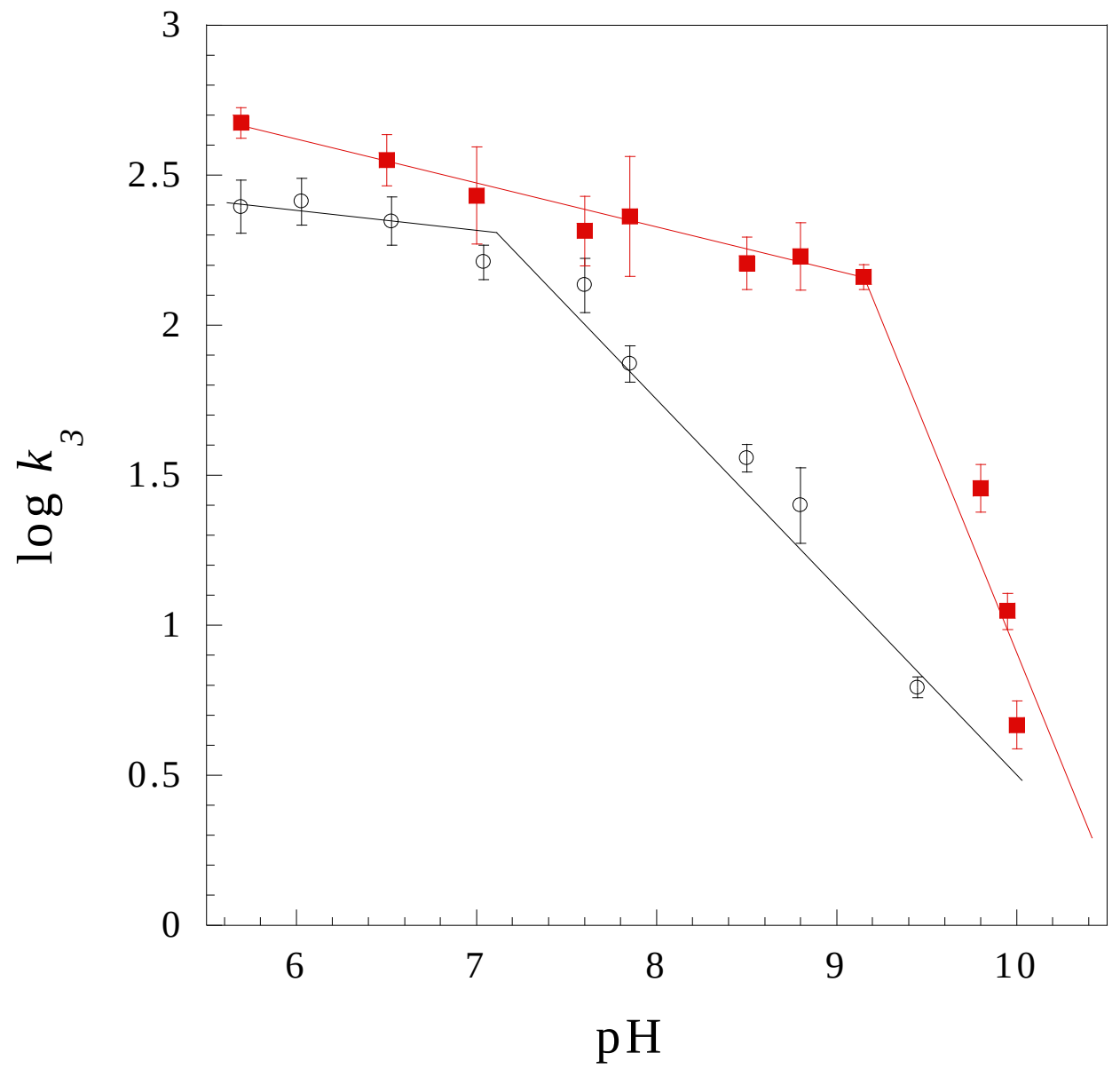

Figure 6

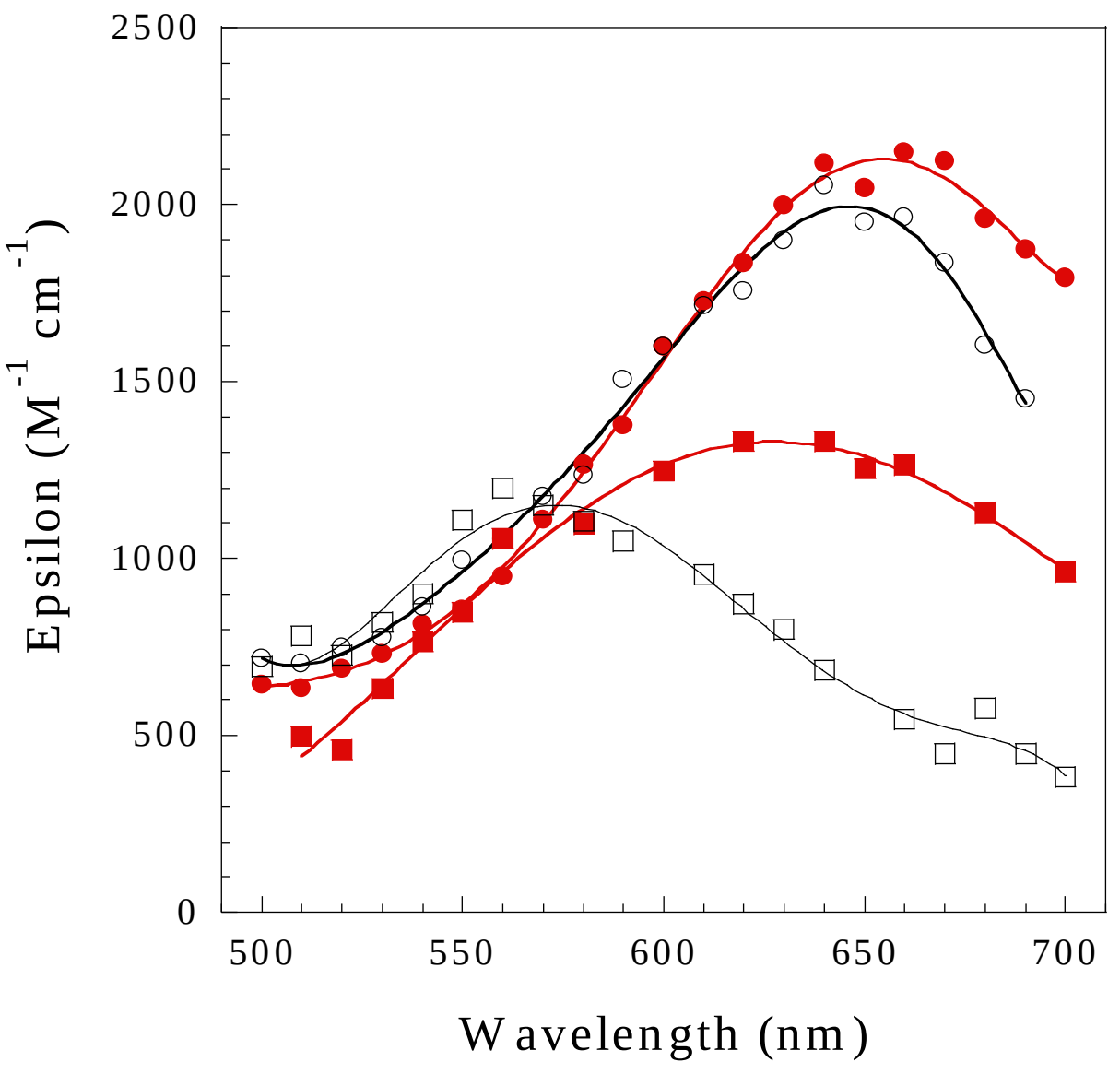

Figure 7

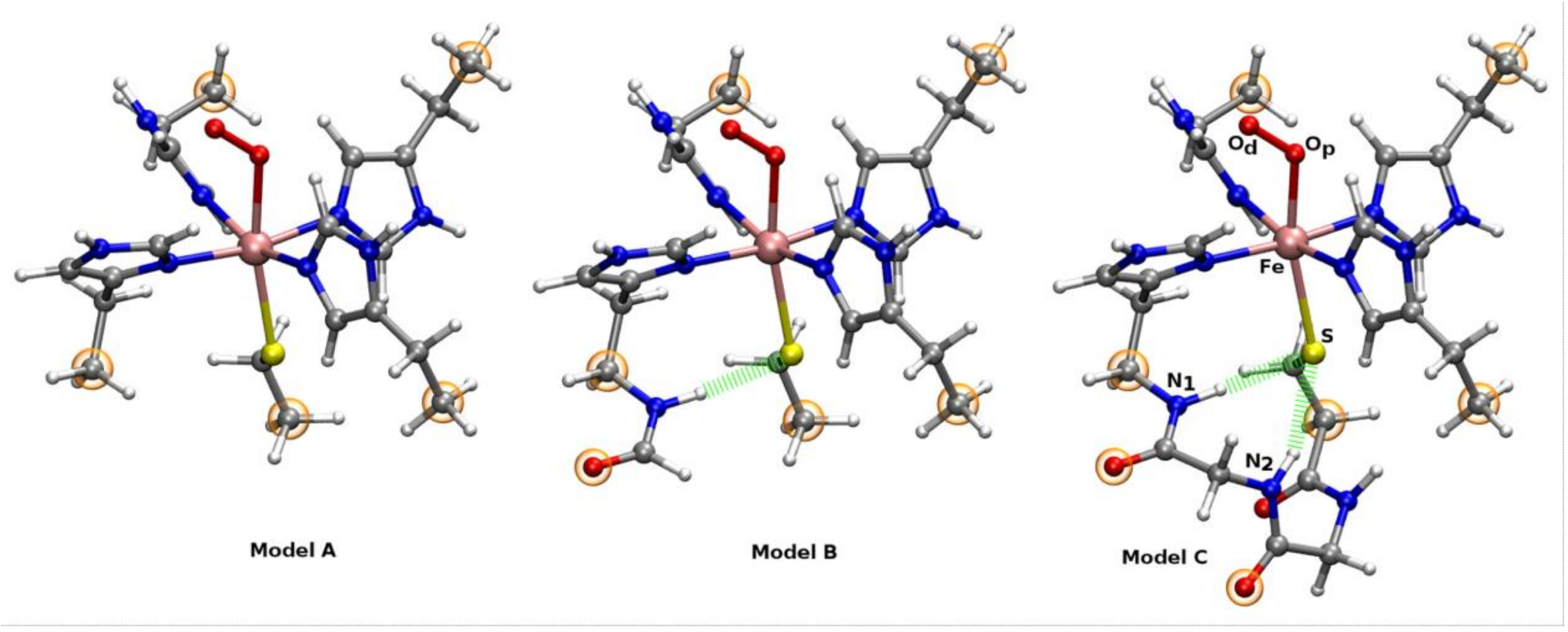

Figure 8